*Terrestrial Planets Comparative Climatology (TPCC) mission concept*

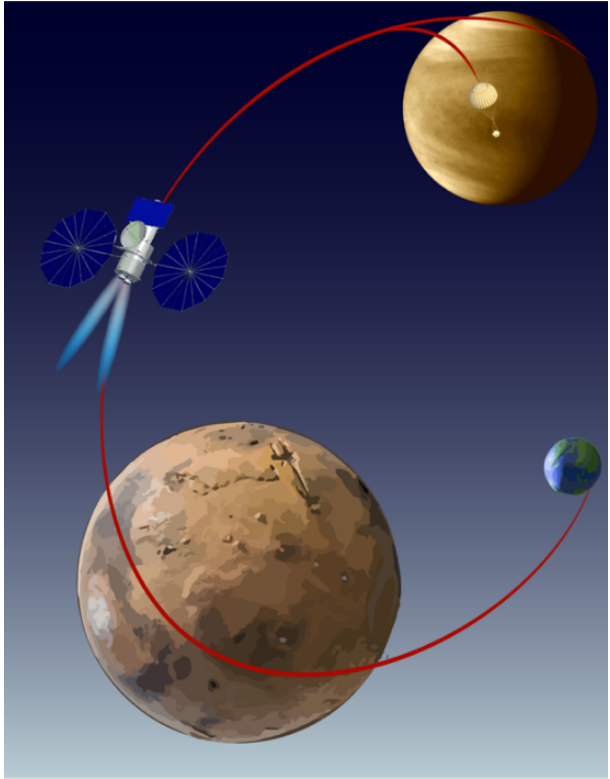

Artist's concept of the Terrestrial Planets Comparative Climatology Mission

**AUTHORS**

Leslie K. Tamppari
Jet Propulsion Laboratory/California Institute of Technology
(818) 653-8348
leslie.tamppari@jpl.nasa.gov

Amanda Brecht[1], Kevin Baines[2], , Brian Drouin[2], Larry Esposito[3], Scott Guzewich[4], Richard Hofer[2], Kandis Lea Jessup[5], Armin Kleinböhl[2], Tibor Kremic[6], Michael Mischna[2], Nicholas Schneider[3], Aymeric Spiga[7]

[1]Ames Research Center
[2]Jet Propulsion Laboratory/California Institute of Technology
[3]University of Colorado, Boulder
[4]NASA Goddard Space Flight Center
[5]Southwest Research Institute
[6]NASA Glenn Research Center
[7]Laboratoire de Météorologie Dynamique

The research was carried out at the Jet Propulsion Laboratory, California Institute of Technology, under a contract with the National Aeronautics and Space Administration (80NM0018D0004). Copyright: © 2020. California Institute of Technology. Government sponsorship acknowledged. CL#20-2520

# 1 Executive Summary

The authors and co-signers of the *Terrestrial Planets Comparative Climatology (TPCC) mission concept* white paper advocate that planetary science in the next decade would greatly benefit from comparatively studying the fundamental behavior of the atmospheres of Venus and Mars, contemporaneously and with the same instrumentation, to capture atmospheric response to the same solar forcing, and with a minimum of instrument-related variability.

NASA's vision is to understand the effects of the Sun on the solar system and the potential for life elsewhere (Obj, 1.1 of [1]). This vision directly motivates the study of terrestrial climate histories and evolution, with the ultimate goal of understanding what factors allow the emergence and sustainability of life/biosphere. It is widely accepted that Venus, Mars, and Earth were formed at the same time and in the same neighborhood of the solar nebula [2, 3], and had water—a necessary ingredient for life. In spite of strong overlap in the atmospheric composition of these planets, their current climatic states are vastly different. While Earth retains water, both Mars and Venus' surfaces are now dry. The manner in which these terrestrial bodies evolved, and whether Mars or Venus currently hosts or previously hosted life, are fundamental mysteries in planetary science. As such, the TPCC mission concept has been developed.

The TPCC guiding science goals are: (1) to understand Venus' evolutionary history compared to Mars and Earth, and (2) to elucidate the driving forces and mechanisms of global circulation at Mars and Venus. These goals address the science question: Why have Mars' and Venus' climates evolved so differently from each other and from Earth's? Furthermore, a greater depth of understanding our own solar system, enables us to understand the variability and likelihood of habitable conditions within extra-solar planetary systems.

We suggest a mission architecture and payload that would determine the noble gas content and isotopic ratios, including the D/H ratio, at Venus with sufficient accuracy to identify the timeline of Venus' climatic evolution for comparison with Mars. In addition, our concept would systematically "follow the energy" at Mars and Venus, in order to establish how solar forcing on the two atmospheric systems leads to the current balance of dynamical, chemical, and radiative processes responsible for each planet's climatic state. This would include acquisition of the first global atmospheric wind profiles on either planet; global, continuous profiles of aerosols, gases, and water vapor at vertical resolutions higher than any previous mission; and regular UV and IR nightglow observations tracking global circulation on both planets.

To increase the fidelity of the planned planetary comparisons, the TPCC mission concept baseline (estimated to be New Frontiers Class) would send a *single* spacecraft to visit one planet and then the next, allowing measurements of each planet with *the same* instrumentation within similar parts of the solar cycle. The concept uses commercial solar-electric propulsion (SEP; e.g., [4]) similar to the Psyche mission [5], plus a probe sent deep into Venus' atmosphere (e.g., LLISSE concept; [6]). Alternatively, "twin" spacecraft (with added probe to the Venus craft), developed at the same time, and launched close in time, would increase the overlap in operation periods and should also be studied. Either architecture is a timely alternative to independently developed and launched spacecraft payloads, which ultimately are more costly and difficult to inter-calibrate.

# 2 The Case for Comparative Climatology

Comparative climatology has been a subject of renewed interest over the last decade, as evidenced by a series of conferences devoted to the subject, resulting in a *Comparative Climatology of the Terrestrial Planets* book [7]. Inter-comparison of terrestrial planet climatology is a key tool for identifying distinctions and similarities in the origin, evolution and future of terrestrial planets—as advocated by the Venus and Mars community advisory groups [8–10]. These comparisons are essential for understanding planetary habitability and the consequences of human activity. These lessons may then be extended to studies of extra-solar terrestrial planets, and their evolution and potential habitability.

Comparison of the current state of Venus, Mars, and Earth suggests distinctions in the evolutionary history of each of these terrestrial planets. Although each planet formed from the same proto-solar nebular elements, the apparent carbon and water histories are vastly different. Venus likely had enough water for a global ocean for billions of years [11], but which is now lost. Over time, Earth's carbon has become integrated within its crust. Mars has a primarily $CO_2$ atmosphere and had liquid water in its past, but today is cold and dry.

Quantitative studies of the isotopic and noble gas composition of each planet's atmosphere can tell us how their paths diverged. Xenon (Xe) is of particular interest for understanding the origin and evolutions of the terrestrial planets. Ongoing investigations of Mars and Earth have provided some detail on this noble gas, but the measurements for Earth and Mars do not match measurements made elsewhere in the solar system and therefore indicate a major climate-changing event early in solar system history [12]. Measurements of Xe and other noble gases, such as neon (Ne) and argon (Ar) at Venus are either absent or of insufficient precision to contextualize Venus' evolution relative to Mars and Earth. Furthermore, previous D/H measurements at Venus are uncertain, yielding uncertainty in processes responsible [13] and the details of initial water delivery cannot be discerned. By improving these measurements, TPCC would elucidate the role of asteroid/primitive body delivery of water, clarify loss processes, and improve and equalize current understanding of Mars and Venus climate histories.

The climates of solar system bodies are driven by energy inputs from the Sun and atmosphere absorption and transport, and are modified by atmospheric aerosols and trace constituents. Climate models need explicit observations to refine and constrain how these processes combine to create the observed climatic state. Model-data comparisons (and interpretations) benefit from common observations, i.e., those obtained in a standardized manner at each target ([7] and chapters therein) under similar conditions. Comparing the behavioral differences in atmospheric response given variation in input the variables of composition, atmospheric pressure and temperature, and solar radiation at Earth, Mars, and Venus allow in-depth study of how the same fundamental physics leads to diverse climatic states. Key processes include radiative forcing, condensation and vaporization, atmospheric dynamics, chemical cycles, atmosphere-surface interactions, and geochemistry [14]. By obtaining contemporaneous measurements at Mars and Venus, under similar and measured solar energy input conditions, TPCC would elucidate plausible paths to the varying climatic states observed in the terrestrial planets.

## 3   New Measurement Needs: Background and Justification

Key new measurements would be made by the TPCC mission (Table 1-1). In particular, global wind and composition measurements will be made and used to explore the atmospheric dynamics and its regions of transition on both Mars and Venus.

No direct measurements of the global atmospheric circulation (winds) have been made on either Mars or Venus—an essential measurement needed for understanding climate forcing (Guzewich *et al.,* white paper: [15]). Tracking the wind at Mars provides a direct measurement of how the atmosphere responds to forcing and how aerosols, vapor, and trace species are transported. Combined with concurrent measurements of temperature and aerosol abundance, a truly complete picture of the atmosphere and modern climate state can be painted. Measurements of surface

**Table 1-1.** Science traceability for a comparative climatology mission to the terrestrial planets.

| Science Goal | Science Objective | Measurements Needed | Proof-of-Concept Instrument | Fundamental Science Questions [8, 9]* |
|---|---|---|---|---|
| Understanding Venus' evolution history for comparison to Mars (and ultimately Earth) | Determine the loss of Venus atmosphere over time to establish viable evolutionary histories, for comparison to Mars | Measure the noble gases, including D/H ratio, in the Venus atmosphere from the top of the atmosphere to the surface | In-situ neutral mass spectrometer on descent probe | • How did the atmosphere of Venus form and evolve? (V1A-AL)<br>• How is Venus releasing its heat now and how is this related to resurfacing and outgassing? (V2B-OG)<br>• How have the interior, surface, and atmosphere interacted as a coupled climate system? (V1B-IS, V3B-CI) |
| | | Map noble gas spatial distribution (Ar, He) in Venus' upper atmosphere and exosphere | EUV spectrometer on orbiter | |
| Understand the atmospheric forcing and response, ant the mechanisms responsible, in the current climate of terrestrial planets | Determine the magnitude and distribution of energy input to the top of the Venus cloud deck to constrain the forcing processes that control upper atmosphere dynamics; constrain the temporal variations in chemical abundances in the Mars atmosphere | Measure profiles of H, $N_2$, N, $CO_2$, CO, $O_2$, O, $SO_2$, SO, $O_3$, $H_2O$ | Mid-UV to EUV spectro-meter (limb, nadir, and stellar occultations), plus sub-mm | • What is the nature of the radiative and dynamical energy balance on Venus that defines the current climate? Specifically, what processes control the atmospheric super-rotation and the atmospheric greenhouse? (V2A-all; V2B-RB, AE, UA)<br>• What are the morphology, chemical makeup and variability of the Venus clouds, what are their roles in the atmospheric dynamical and radiative energy balance, and what is their impact on the Venus climate? (V2B-RB, AE, UA)<br>• How have the interior, surface, and atmosphere interacted as a coupled climate system? (V1B-IS, V3B-CI)<br>• What is the mean, wave, and instantaneous global circulation and what is its role in transport? (M2A1.1)<br>• What is the vertical distribution of water and dust and how are these connected to atmospheric circulation? (M2A1.2)<br>• What are the processes that control the chemical composition of the atmosphere? (M2A3)<br>• What is the spatial distribution of aerosols, neutral species, and ionized species in upper atm (M2A3.1)<br>• What are forcings that control the dynamics and thermal structure of the upper atmosphere (coupling to middle atmosphere)? (M2A4.1)<br>• What are the spatial and temporal variations in the column abundances of species that play important roles in atmospheric chemistry as or transport tracers? (M2A3.1, M2A3.2, M2A4.2) |
| | | Measure atmospheric temperature profiles during daytime and nighttime (Venus: from 65–110 km) | Sub-mm spectrometer or IR limb+nadir sounder | |
| | | Measure nighttime emissions by NO and $O_2$ (singlet delta) | Mid-UV spectrometer, sub-mm ($O_2$ singlet delta), MUV auroral emissions | |
| | Determine the net energy input to the Martian atmosphere to constrain the forcing of dynamics (0–100 km) | Measure day and night aerosol and water vapor abundance profiles | IR limb+nadir sounder, plus sub-mm spectrometer | |
| | | Measure day and night temperature profiles | | |
| | Determine the net energy input to the top of the Venus cloud deck to constrain the forcing processes that control upper atmosphere dynamics | Measure cloud optical depth as a function of height and aerosol properties for $SO_2$, SO, $O_3$, $H_2O$, OCS, HCl at 63 ±3 km altitude | Near-IR spectrometer, IR limb+nadir sounder, sub-mm spectrometer | |
| | | Measure temperature, pressure, upward/downward radiation flux in cloud layer | Temperature/pressure sensor and upward/downward flux radiometer on descent probe | |
| | | Measure temperature profiles in the cloud upper layer on Venus | Sub-mm spectrometer or IR limb sounder | |
| | Measure the net energy to the Martian surface | Measure day and night Martian surface temperature and bi-directional reflection | IR limb+nadir sounder with UV/vis. channel | |
| | Determine the dynamical behavior of the atmospheric transition region on Venus (60–100 km) and the lower and middle Martian atmosphere (0–100 km) to constrain forcing of dynamics | Measure the horizontal wind velocities in these regions | Sub-mm spectrometer and IR limb sounder | |
| | | Measure atmosphere temperature and pressure profiles in these regions and H, O for study of escape | | |
| | | Map noble gas spatial distribution (Ar, He) in Mars' atmosphere, as tracers for transport | EUV spectrometer on orbiter | |

*Specific goal and investigation numbers are denoted within the table as, e.g., MEPAG Goal 2, Objective A, Sub-objective A1, Investigation 1  is denoted as M2A1.1, and VEXAG Goal 1, Objective A, Investigation AL is denoted as V1A-AL.*

energy balance across a wide spectral range—UV, visible, near-IR, and thermal-IR—would drastically improve our understanding of the energy flux into and out of the planet and how the atmosphere moderates that flux. Additionally, trace gas measurements would also be used to assess global climatology of key species of Mars from a polar or near-polar orbit—including isotopologues of oxygen in $CO_2$ (e.g., $^{18}OCO$) and carbon in CO (e.g., $^{12}CO$ and $^{13}CO$). These species are important photochemical tracers for the atmospheric circulation (e.g., [16, 17]); they are also important for understanding atmospheric escape, and in refining our understanding of the chemical cycles of $CO_2$ in the martian atmosphere and how it is maintained over geological time (e.g., [13]). Other photochemically relevant species such as $O_3$, NO, and $NO_2$ would also be measured, providing important information about the water cycle (e.g., $O_3$ is a tracer for water vapor saturation conditions; [18] and the astrobiologically-relevant planetary nitrogen cycles [19].

At Venus, global wind measurements would be used to improve numerical studies of the physical drivers of Venus' enigmatic atmospheric circulation patterns. Unlike Mars, Venus has had only a few prime missions to observe and measure its atmosphere and climate, which have been insufficient to address the objectives proposed here; Pioneer Venus, VeGa, Venus Express, and now Akatsuki. Regions of Venus' atmosphere include the sulfuric acid cloud layers at ~48–70 km, the mesosphere (70–90 km) and upper atmosphere (>90 km). In terms of dynamics, Venus' atmosphere can be split into four zones. At the cloud tops (~70 km), the mean atmospheric motion is dominated by the stable retrograde super-rotating zonal (RSZ) wind that is ~60–80 times faster than the planetary rotation. A meridional Hadley-cell circulation with the westward retrograde super-rotation dominates in the middle atmosphere. At altitudes between ~90 and 120 km the circulation is dominated by subsolar to antisolar (SSAS) winds, resulting from the pressure gradient produced by inhomogeneous heating from solar radiation; with the circulation transitioning from RSZ to SSAS between ~70 and 90 km [20]. Above 120 km, the SSAS is dominant but there is evidence of a residual RSZ wind—suggesting a blending of the circulation patterns. How each of these circulation patterns is produced and maintained is poorly understood, as is the exact mechanisms that govern interactions between the zones. For example, at altitudes >120 km, the blending may be driven by waves (e.g., [21, 22]) and/or ion-neutral drag [23]. Additionally, while on average it seems that the equator-to-65°-latitude region moves in Hadley-cell type fashion, above 65° latitude, a polar vortex linked to solid body rotation prevails [24, 25]. Potential mechanisms for supporting cloud top level super-rotation include thermal tides transporting momentum upwards [26], redistribution of angular momentum via waves and mean circulation [e.g., 27, 28], and topography-induced gravity waves [29]. Yet, it is known that a strong local time-dependent wind shear connects the regions extending from the cloud layers to the upper (>90 km) atmospheric regions. So, it is expected that such strongly variable conditions may easily break the stationary waves, releasing momentum, with a potential impact on the super-rotation— but at what altitude this phenomenon occurs and how it contributes to Venus' dynamic system is a critical unknown. Likewise, the altitudes and physical drivers of the transition from super-rotation to SSAS in the 70–90 km region are not well characterized [30].

On Mars, the forcing of radiatively active water ice clouds has been shown to drive waves and tides in the atmosphere that affect the temperature structure of the Martian atmosphere globally [31]. As clouds are modulated by atmospheric tides [32, 33], there is a feedback between clouds and tides that is not understood to date. The propagation of waves and tidal modes not only influences the temperature in the middle atmosphere but also the density of the upper atmosphere and transports energy from the lower to the upper atmosphere [34]. The transition region between Mars' middle and upper atmosphere (~80–120 km) is still largely unexplored.

Combining the synergistic wind observation and dynamic model studies with planned isotopic measurements would also provide the data needed to better understand the evolution of each atmosphere. On Mars, geologic evidence suggests that the atmosphere used to be considerably more massive and probably also warmer and wetter than today. D/H ratios show that a considerable part of this original atmosphere has been lost to space over geologic time [35], likely through photodissociation of water and escape of atomic H [e.g., 36]. Recent results [37] suggest that middle atmospheric water vapor can affect this process, even reducing the gap between estimates of integrated hydrogen loss and present-day escape rates. Currently, direct water vapor measurements at the key altitudes (60–80 km) are sparse and transport processes from the lower to the upper atmosphere are not well characterized. TPCC would acquire the data need to improve this characterization.

Likewise, TPCC-acquired $O_2$ IR nightglow and NO UV nightglow emission detections would be used to investigate connections between energy, dynamics, and chemistry on Venus and Mars. These emissions result from the recombination of dayside photolysis products, which are dynamically transported to the nightside. Traces of the distribution of these emission with local time allows study of the feedbacks between the dayside and nightside chemistry and atmospheric dynamics relative to the solar input. Combining the nightglow measurements with measurements of temperature, winds, and chemical species would provide a comprehensive global picture of atmospheric dynamics at each planet. Although nightglow emissions were obtained at Venus during Venus Express [e.g., 38], and ongoing observations began at Mars by the MAVEN mission in 2014, nightglow emission datasets obtained at Venus and Mars within the same solar cycle do not exist. TPCC would support the direct comparison of atmospheric response to solar forcing relative to the solar cycle, removing uncertainty associated with solar cycle activity, representing a key advancement in our understanding of the atmospheres of these worlds.

## 4   TPCC Mission Concept Relevance

The TPCC mission guiding science goals are to (1) understand Venus' evolutionary history for comparison with Mars and the Earth, and (2) elucidate the driving forces and mechanisms of global circulation at each planet. These goals directly address cross-cutting themes and priority questions identified in the National Academies CAPS report [39] Table A.1 including: (i) building new worlds (evolution of inner planet atmospheres); (ii) investigating Planetary habitats (Mars/Venus evolution); and (iii) identifying Workings of Solar systems (roles of physics, chemistry, geology, and dynamics in driving planetary atmospheres). Since 1989, no new US-led mission has been dedicated to Venus. Although US-led Mars and internationally-led Venus missions have been active, neither the mission duration nor the payloads of these missions sufficiently overlapped to complete investigation of solar forcing contemporaneous at each planet. And, lacking any in-situ elements for the active Venus missions, detailed study of water history at Venus remains unresolved. The TPCC mission would include contemporaneous study of the atmospheric responses to solar conditions, and would provide the critical remote and in-situ data needed to investigate (1) the origin and diversity of terrestrial planet climates; (2) the relationship between the climate evolution and the origin and evolution of life at each planet; and (3) the processes that control climate on Earth-like planets as prioritized by Vision and Voyages ([40] p. 5-3 to 5-5).

The applicability of the high-fidelity comparative investigations enabled by the TPCC mission to both inner solar system evolution and exoplanet terrestrial analog studies is a key advantage and potentially saves cost compared to sending multiple large spacecraft as envisioned by other mission concepts [e.g., 41, 42]. Additionally, the *TPCC concept addresses 12 VEXAG objectives and 7 MEPAG objectives* included in the past and recently updated Goals and Exploration

Roadmap Documents [8–10]. This concept would also contribute to the comparative climatology set of objectives advocated in the white paper by McGouldrick et al. [43], as well as complete almost all objectives outlined in the white paper by Brecht et al. [44]. Finally, the TPCC mission concept addresses exoplanet exploration questions identified by Showman et al. [45].

## 5   TPCC Mission Concept: Advantages and Details

Because the TPCC mission would study two planetary bodies with *the same* payload, standardized measurements would be obtained at each target. This would decrease the complexity (propagated error) included in the data analysis, facilitating a higher fidelity determination of the impact of the Sun on individual components in each climate system, including radiative feedback, dynamics, cloud formation rates, and atmospheric chemistry. The Venus portion of the TPCC concept is a hybrid of the Multi-Platform Mission Option B and Option C concepts defined in [10], and given the global access to winds, temperature, and aerosols would actually accommodate many of the goals associated with the variable altitude cloud level aerial platform.

The cost drivers identified are: a long mission duration with navigation in/out of orbit at two planets, sophisticated instruments ensuring that the science goals for each planet are met, a descent probe deployed into the Venus atmosphere, and possibly two spacecraft. The technology readiness levels for elements of the TPCC mission are high. The proof-of-concept instruments (Table 1-1) are all TRL 9 with the exception of the sub-mm sounder, which is TRL 5 but could be brought to TRL 6 with very modest cost/schedule, and the descent probe. Despite its powerful ability to provide fundamental climate measurements, sub-mm sounding is a tool that has never been used at Mars and Venus, despite being envisioned in multiple studies and proposals [e.g., 46, 47, 42].

**A MEPAG 1-pager can be found at** https://mepag.jpl.nasa.gov/meetings.cfm?expand=m38
**White Paper #15.**
**Co-Signers to this White Paper can be found at:**
https://docs.google.com/spreadsheets/d/1gJTSmeRvZfOEdhamhLAJo40sY9Z7LoAviDKF2qtDihU/edit?usp=sharing

Pre-Decisional Information – For Planning and Discussion Purposes Only